\documentclass{css2018}

\usepackage{epsfig}

\newcommand{\PACS}{\MSC}

\title{Problems with the dark matter and dark energy hypotheses, and alternative ideas}

\author{Mart\'\i n L\'opez-Corredoira$^{1,2}$\\
\vskip 2mm {\small
$^1$
Instituto de Astrof\'\i sica de Canarias\\
C/.V\'\i a L\'actea, s/n, E-38205 La Laguna (Tenerife), Spain\\
martinlc@iac.es \\
$^2$
Department of Astrophysics, University of La Laguna\\
E-38206 La Laguna (Tenerife), Spain \\
}}

\abstract{Two exotic elements have been introduced into the standard cosmological model: non-baryonic dark matter and dark energy. The
success in converting a hypothesis into a solid theory depends strongly on whether we are able to solve the problems in explaining observations with
these dark elements and whether the solutions of these problems are unique within the standard paradigm without
recourse to alternative scenarios. We have not achieved that success yet because of numerous inconsistencies, mainly on galactic scales, the non-detection so far of candidate particles for dark matter, and the existence of many alternative hypotheses that might substitute the standard picture to explain
the cosmological observations. A review of some ideas and facts is given here.}

  % PACS numbers
\begin{document}

\maketitle

\section{History of the idea of Dark Matter}

The existence of dark or invisible matter detectable through its gravitational influence has been known by astronomers for a long time now \cite{Ber16}. Bessel \cite{Bes44} in 1844 argued that the observed proper motions of the stars Sirius and Procyon could be explained only in terms of the presence of faint companion stars. In 1846, Le Verrier and Adams independently predicted the existence of Neptune based on calculations of the anomalous motions of Uranus. Le Verrier later proposed the existence of the planet Vulcan to explain anomalies in the orbit of Mercury, but he failed this time because the solution was not invisible matter but a change of gravitational laws, as was solved years later by Einstein with General Relativity. The dynamical analysis of dark matter in form of faint stars in the Milky Way using the motion of stars was carried out by Lord Kelvin in 1904, Poincar\'e in 1906, \"Opik in 1915, Kapteyn in 1922, Jeans in 1922, Lindblad in 1926, and Oort in 1932 with different results \cite{Ber16}.

With regard to extragalactic astronomy, Zwicky's \cite{Zwi33} 1933 paper on dark matter in rich clusters applied the virial theorem to these data and found
a mass-to-light ratio of $\sim$60 in solar units (rescaled to the
present-day value of the Hubble constant).
In 1959 Kahn \& Woltjer \cite{Kah59} determined the mass of the Local Group
and obtained a mass-to-light ratio of 43 in solar units.
In the 1950s, Page \cite{Pag52,Pag60} also found that pairs of elliptical galaxies had a mass-to-light ratio of 66 in solar units. This showed that such binaries must have massive envelopes or be embedded in a massive common envelope. Similar
results were obtained in the 1950s from 26 binary galaxies by Holmberg \cite{Hol54}.
In 1939 Babcock \cite{Bab39} first showed the need for dark matter for an individual galaxy by measuring the rotation curve of the outer regions of M31 out to 100 arcminutes ($\approx 20$ kpc) from its center.
However, the majority of astronomers did not become convinced of the need for dark
matter halos in galaxies until the publication of theoretical papers in the 1970s, 
such as the one on the stability of galactic disks
by Ostriker \& Peebles \cite{Ost73}. Later, rotation curves in the radio by Albert Bosma \cite{Bos78} and in the visible by Vera Rubin, Kent Ford, and Nortbert Thonnard \cite{Rub80} easily convinced the community.
This shows the typical mentality of astrophysicists:
accepting facts only when there is a theory to support them with an explanation, a not-so-empirical approach that dominates the development of cosmology.

Cosmology has indeed played a very important role in the idea of dark matter on galactic scales. The first predictions based on Cosmic Microwave Background Radiation (CMBR) anisotropies were wrong. It was predicted in the 1960s 
that $\Delta T/T$ should be one part in a hundred or a thousand  \cite{Sac67}; however, 
fluctuations with this amplitude could not be found from observations in the 1970s. In order to solve this problem, non-baryonic dark matter was introduced {\it ad hoc} and was thought to be composed of certain mysterious particles different from known matter. 
In a short time, the connection between particle physics and the missing mass problem in galaxies arose. Many astrophysicists considered dark matter halos
surrounding galaxies and galaxy clusters possibly to consist of a gas of non-baryonic particles rather than faint stars or other astrophysical objects. 
This was a happy idea without any proof; there is no proof that directly connects the problem of the amplitude of CMBR anisotropies with the rotation curves of galaxies or the missing mass in clusters, but the idea was pushed by leading
cosmologists, who made the idea fashionable among the rest of the astrophysical community.

Part of the success of these non-baryonic dark matter scenarios in the halos of the galaxies was due to the good agreement of simulations of large scale structure
with the observed distributions of galaxies. At first, in the 1980s, with the attempt to fit the data using hot dark matter composed of neutrinos, the simulations showed that very large structures should be formed first and only later
go on to form galaxy-sized halos through fragmentation, which did not match the
observations \cite{Whi83}, whereas cold dark matter (CDM) models were more succesful, at least on large scales ($>1$ Mpc).

This tendency towards selling a prediction of failure as a success for a model via
the {\it ad hoc} introduction of some convenient form of unknown dark
matter still prevails. An instance of this predilection is the introduction in 2018
of some peculiar form of dark matter\cite{Bar18} in order to cool the gas at
$z\approx 18$ and solving the discrepancies in the measurements of 21 cm line 
amplitude with respect to the {\it a priori} predictions\cite{Bow18}.

\section{Dark matter and inconsistencies of the theory at 
galactic scales}
  
That there is some dark matter, either baryonic or non-baryonic, is clear, 
but how much, and what is its nature?
The success of the standard model in converting a hypothesis into a solid theory
depends strongly on the answer to these open questions.
Stellar and cold gas in galaxies sum to baryonic matter content that is 8$^{+4}_{-5}$\%
of the total amount of the predicted Big Bang baryonic matter \cite{Bel03}. 
Where is the rest of the baryonic material?
What is the nature of the putative non-baryonic dark matter required
to achieve the current value of $\Omega _m\approx 0.3$?
 
Current CDM models predict the existence 
of dark matter haloes for each
galaxy whose density profile falls approximately as $r^{-2}$, although the
original idea \cite{Whi78} concerning hierarchical structures with
CDM, which gave birth to the present models, was that the dark matter was distributed 
without internal substructure, more like a halo with galaxies than 
galaxies with a halo \cite{Bat00}, something similar to the scenario
in Refs.  \cite{Lop99,Lop02b}. 
 
Some authors have been led to question the very existence 
of this dark matter on galactic scales since its evidence is
weak \cite{Bat00,McG00,Eva01,Tas02} and the predictions do not fit the
observations: 
CDM has a ``small scale crisis'' since there are some features of the galaxies that
are very different from the predictions of the cosmological model. Nonetheless, many
researchers are eagerly trying to find solutions that make data and model compatible, 
assuming a priori that the model ``must be'' correct. Some of the problems are the following.
 
There is a problem with an observed lower density of the halo in the inner
galaxy than predicted. $\Lambda$CDM (CDM including
a $\Lambda $ term for the cosmological constant; see \S \ref{.cosmconst}) predicts 
halo mass profiles with cuspy cores 
and low outer density, while lensing and dynamical observations 
indicate a central core of constant density and a flattish high dark 
mass density outer profile \cite{Per08}.
The possible solutions of core-cusp 
problem without abandoning the standard model are: bar-halo friction, which
reduces the density of the halo in the inner galaxy \cite{Sel06};
haloes around galaxies may have undergone a compression by the  
stellar disc \cite{Gne04} or/and suffered from the effects of baryonic physics \cite{DiC14}. 
 
Another problem is that the predicted angular momentum is much less than the observed one.
Binney et al. \cite{Bin01} claim that the problem of an excess of predicted 
dark matter within the optical bodies and the fact that the observed 
discs are much larger than expected can be solved if a considerable mass of low angular momentum baryons is ejected (massive galactic outflows) and the discs are formed 
later from the high angular momentum baryons which fell in the galaxy.
The conspiracy problem is also solved if the ejection begins only 
once $M_{\rm baryons}(r)\sim M_{\rm dark\ matter}(r)$.
Another solution within the standard cosmological model 
for the angular momentum problem is the tidal interaction of objects populating the primordial voids together with the Coriolis force due to void rotation \cite{Cas15}.

Another fact that could cast doubt upon the existence of very massive halos of dark matter is that strong bars rotating in dense halos should generally slow 
down as they lose angular momentum to the halo through dynamical friction \cite{Deb00}, whereas the observed pattern speed of galactic bars indicates that 
almost all of them rotate quite fast \cite{Agu15}. There should be a net transference of angular momentum from bars to halos, although friction can be 
avoided under some special conditions \cite{Sel06b}.
 
The enclosed dynamical mass-to-light ratio increases with decreasing galaxy luminosity and surface brightness, which is not predicted by dark matter scenarios \cite{McG14}.
 
Galaxies dominate the halo with little 
substructure whereas the model predicts that galaxies should be scaled versions 
of galaxy clusters with abundant substructure  \cite{DOn04,Kro10}.
Moreover, $\Lambda$CDM simulations predict that the majority of the most massive subhalos of the Milky Way are too dense to host any of its bright satellites 
($L_V>10^5$ L$_\odot $) \cite{Boy11}.
Also, the distribution of satellites is 
in a plane, incompatible with $\Lambda$CDM \cite{Kro10,Kro05,Paw13}. Kroupa \cite{Kro12} says that these are arguments against the 
standard model in which one cannot make the typical rebuff of incompleteness of knowledge of baryonic physics. Furthermore, there is a correlation between bulge mass and the number of luminous satellites in tidal streams \cite{Kro10,Lop16} that is not predicted by the standard model, and it is predicted by models of modified gravity without dark matter. The disc of satellites 
and bulge-satellite correlation suggest that dissipational events forming bulges 
are related to the processes forming phase-space correlated satellite populations. 
These events are well known to occur, since in galaxy encounters energy and angular 
momentum are expelled in the form of tidal tails, which can fragment to form 
populations of tidal-dwarf galaxies and associated star clusters. If Local 
Group satellite galaxies are to be interpreted as Tidal Dwarf galaxies then the substructure predictions of the standard cosmological model are internally in conflict \cite{Kro10}. 

Perhaps, that most severe caveat to retain the hypothesis of dark matter is that, after a long time looking for it, it has not yet been found, although non-discovery does not mean that it does not exist.
Microlensing surveys \cite{Las00,Tis07} constrain the mass of the halo in our Galaxy
in the form of dim stars and brown dwarfs to be much less than that
necessary for dark matter halos.
In any case, as already mentioned, the primordial nucleosynthesis model constrains baryonic matter to be around 10\% of the total mass \cite{Bel03}, so these objects could not be compatible with the preferred cosmological model.
Some observations are inconsistent with the dominant dark matter component
being dissipationless \cite{Moo94}. Neither massive black hole halos  \cite{Moo93} nor intermediate-mass primordial black holes \cite{Med17} provide a consistent scenario. The nature of dark matter has been investigated and there are no suitable candidates among
astrophysical objects.

\section{Dark matter particles} 

The other possibility is that dark matter is not concentrated in any kind of 
astrophysical object but in a gas of exotic non-baryonic particles.
There are three possible types of candidates \cite{Ber16}: 1) particles predicted by the supersymmetry hypothesis, which are electrically neutral and not strongly interacting, including superpartners of neutrinos, photons, Z bosons, Higgs bosons, gravitons, and others (neutralinos have been the most recently studied candidates in the
last decades); 2) axions, typically with masses between $10^{-6}$ and
$10^{-4}$ eV, predicted to resolve certain problems in quantum chromodynamics; and 3) Weakly Interacting Massive Particles (WIMPs), which are those particles that interact through the weak force.

The latest attempts to search for exotic particles have also finished without success.
Technologies used to directly detect a dark matter particle have failed to obtain any positive result \cite{Mar16,Liu17}.
Attempts have also been made to detect neutralinos with the MAGIC and HESS Cerenkov telescope systems for
very high energy gamma rays through their Cherenkov radiation, but so far without success 
and only emission associated with the Galaxy has been found \cite{Aha06}.
Dwarf galaxies are expected to have high ratios of dark matter and
low gamma ray emission due to other astrophysical processes so the search is 
focussed on these galaxies, but without positive results.
As usual, the scientists involved in these projects attribute their failure
of detection to the inability of the detectors to reach the necessary lower cross section of the interaction, or to tbe possibility that they may
be 3--4 orders of magnitude below the possible flux of gamma rays emitted 
by dark matter \cite{San09}, and ask for more funding to continue to feed their illusions: a never-ending story. As pointed out by David Merritt \cite{Mer17}, this will never constitute a falsification of the CDM model because although success of detection will confirm the standard paradigm, non-detection is not used to discard it.

\section{Scenarios without non-baryonic cold dark matter}

Note also that some other dynamical problems  in which dark matter has been claimed as necessary can indeed be solved without dark matter: galactic stability  \cite{Too81}
or warp creation \cite{Lop02b}, for instance. Rotation curves in spiral galaxies
can be explained without non-baryonic dark matter with magnetic fields \cite{Bat00}, or modified gravity  \cite{San02}, or baryonic dark matter in the outer disc \cite{Fen15}
or non-circular orbits in the outer disc \cite{Ben17}. 
Velocities in galaxy pairs and satellites 
might also measure the mass of the intergalactic 
medium filling the space between the members of the 
pairs \cite{Lop99,Lop02b} rather than the mass of dark haloes
associated with the galaxies. 

The most popular alternative to dark matter is the modification of gravity laws proposed in MOND (Modified Newtonian Dynamics;  \cite{San15}), which
modifies the Newtonian law for accelerations lower than $1\times 10^{-10}$ m/s$^2$. This was in principle a phenomenological approach. It was attempted to incorporate elements that make it compatible with more general gravitation theories. The AQUAdratic Lagrangian theory 
(AUQAL)  \cite{Bek84} expanded MOND to preserve the conservation of momentum, angular momentum, and energy, and follow the weak equivalence principle. Later, a relativistic gravitation theory of MOND would be developed under the name  Tensor-Vector-Scalar (TeVeS) \cite{Bek04},
which also tried to provide consistency with certain cosmological observations, including gravitational lensing.
However, the successes of MOND and its relativistic version are mostly limited
to galactic scales and cannot compete with $\Lambda $CDM to explain the
large-scale structure and other cosmological predictions.
Moreover, a search was made for evidence of the MOND statement in a terrestrial laboratory: a sensitive torsion balance was employed to measure small accelerations due to gravity, and no deviations from
the predictions of Newton's law were found down to $1\times 10^{-12}$ m/s$^2$
 \cite{Lit14}. Therefore, unless these experiments are wrong, or we interpret the transition regime acceleration of $1\times 10^{-10}$ m/s$^2$ in terms of total absolute acceleration (including the acceleration of the Earth, Sun, etc.) rather than the relative one, MOND/TeVes is falsified by this experiment.

There are also proposals that the dark matter necessary to solve many problems may be baryonic: positively charged, baryonic (protons and helium nuclei) particles \cite{Dre05}, which 
are massive and weakly interacting, but only when moving at relativistic velocities; 
simple composite systems that include nucleons but 
are still bound together by comparable electric and magnetic forces \cite{May12}, making up a three-body system ``tresinos'' or four -body system ``quatrinos''; antiparticles which have negative gravitational charge \cite{Haj14}, etc.

In my opinion, the problem of `dark matter' is not only one problem but many different problems within astrophysics that might have different solutions. The idea that the same
non-baryonic dark matter necessary to explain the low anisotropies in the CMBR is going to solve the large-scale structure distribution, the lack of visible matter in clusters, the dispersion of velocities of their galaxies, the measurements of gravitational lensing, the rotation curves, etc., is a happy fantasy that has dominated astrophysics for the last 40 years. It would be wonderful if we also get a happy ending with the discovery of the particles
of dark matter that constitute the dark halos of galaxies, but, in absence of that outcome, maybe
it would be prudent to bet on a combination of different elements to explain the entire
set of unexplained phenomena: possibly some baryonic dark matter in some cases, possibly
a modification of gravity is part of the explanation for a wide set of events, and 
maybe cold dark matter dominates some phenomena and hot dark matter other phenomena. Certainly, a unified picture of a unique non-baryonic type of cold dark matter to explain
everything would be a simpler and more elegant hypothesis; the question, however, is not
one of simplicity but one of ascertaining how reality is, whether simple or complex.
 
\section{Dark energy and the cosmological constant or quintessence}
\label{.cosmconst}
  
The question of the cosmological constant  to maintain a static universe  \cite{Pad03} was considered Einstein's biggest blunder, 
and it was introduced by Lema\^itre  \cite{Lem34} in his equations for the evolution of the expanding universe. Indeed, it is equivalent to positing an attractive  gravitational acceleration $a(r)=-\frac{GM}{r^2} + Br$, already proposed by Newton for $B<0$, but with $B>0$ instead \cite{Kom11}. It is not usual physics but an exotic suggestion, since the usual thermodynamics for fluids with positive heat capacity and positive compressibility is not appliable to dark energy with negative pressure \cite{Bar15}.

Twenty-five years ago, most cosmologists did not favour the scenarios dominated by the cosmological constant \cite{Fuk91}.
In the eighties, the cosmological constant was many times disregarded as an 
unnecessary encumbrance, or its value was set at zero \cite{Lon86},
and all the observations gave a null or almost null value.
However, since other problems in cosmology have risen, many cosmologists
at the beginning of the '90s realized that an $\Omega _\Lambda $ ranging from 0.70 to 0.80 could solve many problems in CDM cosmology \cite{Efs90}. Years later, evidence for such
a value of the cosmological constant began to arrive. A brilliant prediction
or a prejudice which conditions the actual measurements?

All present claims about the existence of dark energy have 
measured $\Omega _\Lambda $ through its dependence on the luminosity distance 
vs. redshift dependence  \cite{Dur11}. In the mid-1990s the position of the first peak in the power spectrum of the CMBR was determined to be at $\ell \approx 200$. White et al.\ in 1996  \cite{Whi96} realized that the preferred standard model at that time (an open universe with $\Omega =\Omega _m\approx 0.2$ and without dark energy) did not fit the observations, so that they needed a larger $\Omega $.
Between 1997 and 2000 a change of mentality in standard cosmology occurred.
This was one of the elements, together with Type Ia Supernovae  (SN Ia) observations and the age problem of the universe, that would encourage cosmologists to include a new ad hoc element: dark energy.

One measurement of the cosmological constant comes nowadays from
supernovae, 
whose fainter-than-expected luminosity in distant galaxies can be explained
with the introduction of the cosmological
constant. It was criticized as being due possibly to
intergalactic dust \cite{Agu00,Goo02,Mil15}. The presence of grey dust is
not necessarily inconsistent with the measure of a supernova at $z=1.7$ (SN 1997ff) \cite{Goo02}.
Dimming by dust along the line of sight, predominantly in the
host galaxy of the SN explosion, is one of the main
sources of systematic uncertainties \cite{Kno03}.
Also, there was an underestimate of the effects of host galaxy extinction: a 
factor which may contribute to apparent faintness of 
high-$z$ supernovae is the evolution of the host galaxy extinction with $z$  \cite{Row02}; 
therefore, with a consistent treatment of host galaxy
extinction and the elimination of supernovae not observed before maximum, 
the evidence for a positive $\Lambda $ is not very significant.
Fitting the corrected luminosity distances (corrected for internal extinctions)
with cosmological models Balazs et al. \cite{Bal06} concluded that 
the SNIa data alone did not exclude the possibility of the $\Lambda=0$ 
solution.
 
SNe Ia also possibly have a metallicity dependence and
this would imply that the evidence for a non-zero cosmological constant 
from the SNIa Hubble Diagram may be subject to corrections for metallicity
that are as big as the effects of cosmology \cite{Sha01}. 
The old supernovae might be intrinsically fainter than the local ones,
and the cosmological constant would not be needed \cite{Dom00}.
As a matter of fact, some cases, 
such as SNLS-03D3bb, have an exceptionally high luminosity \cite{How06}. 
Claims have been made about the possible existence of 
two classes of Normal-Bright SNe Ia \cite{Qui07}.
If there is a systematic evolution in the metallicity of SN Ia progenitors, 
this could affect the determination of cosmological parameters. 
This metallicity effect could be 
substantially larger than has been estimated previously and could
quantitatively evaluate the importance of metallicity evolution 
for determining cosmological parameters \cite{Pod06}. 
In principle, a moderate and plausible amount of metallicity 
evolution could mimic a $\Lambda$-dominated, a flat universe in an open, 
$\Lambda $-free universe. However, the effect of metallicity evolution 
appears not to be large enough to explain the high-$z$ SNIa data in a 
flat universe, for which there is strong independent evidence, 
without a cosmological constant. 
 
Furthermore, our limited knowledge of the SN properties in the U-band
has been identified as another main source of uncertainty in the
determination of cosmological parameters \cite{Kno03}.
And the standard technique with SNe Ia consists in using spectroscopic
templates, built by averaging spectra of well observed (mostly nearby)
SNe Ia. Thus, the uncertainty in K-corrections depends primarily
on the spectroscopic diversity of SNe Ia.
 
Even if we accept the present-day SN Ia analyses as correct and without any
bias or selection effect,
other cosmologies may
explain the apparent cosmic acceleration of SNe Ia without introducing a cosmological 
constant into the standard Einstein field equation, thus negating the necessity for 
the existence of dark energy \cite{Shu10}. There are four distinguishing features of these 
models: 1) the speed of light and the gravitational ``constant'' are not constant, 
but vary with the evolution of the universe, 2) time has no beginning and no 
end, 3) the spatial section of the universe is a 3-sphere, and 4) the universe 
experiences phases of both acceleration and deceleration. 
An inhomogeneous isotropic universe described by a
Lema\^itre--Tolman--Bondi solution of Einstein's fields equations can also provide
a positive acceleration of the expansion without dark energy \cite{Rom07}.
Quasi-Steady-State theory predicts a decelerating 
universe at the present era, it explains successfully the 
recent SNe Ia observations \cite{Vis07}.
Carmeli's cosmology fits data for an accelerating and decelerating universe 
without dark matter or dark energy \cite{Oli06}.
Thompson \cite{Tho12b} used available measurement for the constrainst on the variation
the proton to mass electron with redshift, and with $\frac{\Delta \alpha }{\alpha }
=7\times 10^{-6}$ he finds that almost all of the dark energy models 
using the commonly expected values or parameters are excluded.
A  static universe can also fit the supernovae data without dark energy \cite{Sor09,Ler09,Lop10a,Far10,Mar13}.
 
There are other sources of $\Omega _\Lambda $ measurement such as the anisotropies
of the CMBR, but they are not free of inaccuracies owing to contamination and anomalies found in it  \cite{Lop07,Sch16}.
In the last two decades, many proofs have been presented to the community 
to convince us that the definitive cosmology has $\Omega _\Lambda \approx 0.7$,
which is surprising taking into account that in the rest of the history
of the observational cosmology proofs have been presented for $\Omega _\Lambda \approx
0$. Furthermore, recent tests indicate that other values are available in the
literature. For instance, from the test angular size vs.\ redshift for
ultracompact radio sources, it is obtained that $\Lambda $ is negative \cite{Jac97}. 
Using the brightest galaxies in clusters, the fit in the
Hubble diagram is compatible with a non-accelerated universe  instead of $\Omega _\Lambda =0.7$  \cite{Vau03,And06}. Concordance models produce far more high redshift massive clusters
than observed in all existing X-ray surveys \cite{Bla06}.
 
The actual values of $\Omega _\Lambda $ have some consistency problem 
in the standard scenario of the inflationary Big Bang. 
The cosmological constant predicted by quantum field theory has
a value much larger than those derived from observational cosmology.
This is because the vacuum energy in quantum field
theory takes the form of the cosmological constant in Einstein's equations.
If inflation took place at the Grand Unified Theory
epoch, the present value would be too low by a factor $\sim 10^{-108}$, and
if the inflation took place at the quantum gravity epoch, the above factor would be lower still at $\sim$$10^{-120}$  \cite{Wei89}.
The intrinsic absence of pressure in the ``Big Bang Model'' also
rules out the concept of ``Dark Energy'', according to some opinions \cite{Mit11}.
 
Furthermore, the standard model has some surprising coincidences.
There is the coincidence that now the deceleration of the Hubble flow
is compensated by the acceleration of the dark energy;
the average acceleration throughout the history of the universe is almost null \cite{Mel12}. 
Again, everything is far from being properly understood.

%\begin{figure}[h]
%\begin{center}
%\includegraphics[width=9cm]{figure}
%\caption{\label{fig:1} This figure was created in Linux by \texttt{xfig}.}
%\end{center}
%\end{figure}

%
%\begin{table}[ht]
%\begin{center}
%\begin{tabular}
%[c]{lrrrrr}
%\hline
%\#proc                 &  64    & 128   & 256   & 512   &  1024 \\
%\hline
%\multicolumn{6}{c}{ case 1 }\\
%\hline
%set-up (sec)           &  61.0  & 37.7  & 25.7  & 23.2  &  39.5 \\
%iter (sec)             &  22.3  & 19.9  & 27.8  & 44.9  &  97.5 \\
%\hline
%\multicolumn{6}{c}{ case 2 }\\
%\hline
%set-up (sec)           &  49.5  & 29.0  & 18.4  & 12.6  &  11.0 \\
%iter (sec)             &  28.5  & 22.6  & 16.7  & 14.7  &  13.2 \\
%\hline
%\end{tabular}
%\end{center}
%\caption{\label{tab:1} Strong scaling for different cases.}
%\end{table}

\section*{Acknowledgements}
Thanks are given to the language editor Terence J. Mahoney (IAC, Tenerife, Spain) for proof-reading of the text. The author was supported by the grant AYA2015-66506-P of the Spanish Ministry of Economy and Competitiveness (MINECO).

\end{document}